\documentclass[twocolumn]{aastex631}

\begin{document}

\title{Little time for oscillation: Fast disruption of the Radcliffe Wave by Galactic motions}

\author[0000-0003-3144-1952]{Guang-Xing Li}
\altaffiliation{E-mail: gxli@ynu.edu.cn (G-XL); bchen@ynu.edu.cn (B-QC) }
\affiliation{ South-Western Institute for Astronomy Research, Yunnan University, Chenggong District, Kunming 650091, P. R. China}


\author[0000-0002-1463-9732]{Ji-Xuan Zhou}

\affiliation{ South-Western Institute for Astronomy Research, Yunnan University, Chenggong District, Kunming 650091, P. R. China}
\affiliation{School of Physics and astronomy, Cardiff University, Queen’s Buildings, The Parade, Cardiff, CF24 3AA, UK}

\author[0000-0003-2472-4903]{Bingqiu Chen}
\affiliation{ South-Western Institute for Astronomy Research, Yunnan University, Chenggong District, Kunming 650091, P. R. China}


\begin{abstract}
     The Radcliffe wave \cite{2020Natur.578..237A} is a 2.7 kpc long, 100 pc wide-like structure in the Galactic disk with a wave-like velocity structure \cite{2022MNRAS.517L.102L,2024arXiv240212596K}. A referent Nature paper \cite{2024arXiv240212596K} treated the Wave as a solid body in the disk plane, modeled its oscillation along the vertical direction, and derived the local Galactic mass distribution from the oscillation pattern. In reality,
     Galactic shear can stretch gas through differential rotation, whereas gas clouds experience epicyclic motions. 
     We simulate the 3D evolution of the local interstellar gas and find shear and encyclic motion stretches the Radcliffe wave to almost twice its current length at the timescale of 45 Myr, within which only half a cycle of the proposed vertical oscillation occurs. The simulation also reveals the formation of new filaments and filament-filament mergers. Treating the Radcliffe wave as a solid body in the Galactic disk and an oscillating structure in the vertical direction is thus an oversimplification. Our data-driven simulation reveals the 3D evolution of the local interstellar gas with several processes at play, strengthening the role of the Solar Neighborhood as a unique test ground for theories of interstellar gas evolution.
\end{abstract}



\section{ Radcliffe wave in context}
The Radcliffe wave \cite{2020Natur.578..237A} is a long, coherent gas structure that contains several star-forming regions. Galactic-scale gas filaments as such were discovered almost a decade ago \cite{2013A&A...559A..34L}, and their evolutions are affected by other processes from the Galaxy, such as shear and the feedback processes related to the formation of stars in  star-forming regions. In our previous paper, we discovered the coherent velocity structure of the wave using motions from young stars associated with the Wave \cite{2022MNRAS.517L.102L}.  In a recent Nature paper \cite{2024arXiv240212596K}, the authors used line-of-sight velocity for $^{12}$CO and 3D velocities of young stellar clusters to show that (1) ``the Radcliffe Wave is oscillating through the Galactic plane'' and (2)  they can ``derive its motion independently of the local Galactic mass distribution, and
directly measure local properties of the Galactic potential as well as the Sun's vertical oscillation period''.

\section{ Motion of gas in the Galactic disk}
Structures in the Milky Way disk are subject to influences from the Galactic potential, whose effects can be summarized as follows \cite{1987gady.book.....B}:
\begin{enumerate}
  \item {\bf Shear:} The flat rotations curve of the Milky Way and other disk galaxies leads to differences in angular velocity $\omega = v / r$ at different radii. This shearing motion causes patterns on galaxies to wind up \cite{1996ima..book.....C}. 
  The shear time is
  \begin{equation}
    t_{\rm shear} = 2 A \approx  30  \, {\rm Myr} \;,
   \end{equation}
   where $A \approx 15\; \rm km\, s^{-1}\,kpc^{-1}$ is the Oort constant.
   
  \item {\bf Vertical oscillation:} Along the vertical direction, the motions of clouds are affected by gravity from the mostly the visible matter. 
   The timescale for vertical oscillation can be estimated using the equation \footnote{ Where $a_z = 4 \pi G\rho_0 z$, $\omega_{\rm oscillation} = \sqrt{4 \pi G \rho_0}$.}
  \begin{equation} 
     t_{\rm oscillation}  = \sqrt{\pi / (G\rho_0)} \;,
   \end{equation}
  where, assuming $\rho_0 =\, 0.084\, M_\odot\rm pc^{-3}$ \cite{2015ApJ...814...13M}, we have $t_{\rm oscillation} \approx$ 90 Myr, consistent with previous estimates. 

  \item {\bf Epicyclic motion:} Motion of particles around the guiding centers. 

\end{enumerate}

The Nature paper \cite{2024arXiv240212596K} only considered the motion along the vertical direction. In the disk plane, the Wave was assumed to be a solid body where internal motions caused by the Galactic potential are neglected. 
Since $t_{\rm shear} < t_{\rm \rm oscillation}$, the effect of shear should not be neglected. In our previous papers \cite{2022MNRAS.516L..35L,2024MNRAS.529.1091Z}, we have measured the 3D velocities of a sample of clouds by combining  $^{12}$CO  observations of the radial velocity of the gas with proper motions of stars associated with the gas and modeled the 2D motion of the gas clouds in the Galactic disk.
 In this letter, we added the vertical dimension and model the 3D evolutions of the local interstellar gas.

 We follow the approach of the previous paper  \cite{2022MNRAS.516L..35L}, with some further improvements:  the simulation is performed in the Local Co-rotating Frame \cite{2022MNRAS.516L..35L} where a term representing the Colaris force is added \cite{1987gady.book.....B}. The motion along the $z$ direction is modeled as harmonic oscillators of $t_{\rm oscillation} = 90\;\rm Myr$. We did not include the aharmonic corrections as in the Nature paper, since our focus is to understand the effect of the horizontal motions on the overall picture. The aharmonic effects are minor as the corrections to the vertical oscillation time are only a few percent \cite{2020Natur.578..237A}.  
The results are plotted in Fig. \ref{fig:1}.  An animation is available at \url{see-attachment-link-to-be-inserted-later}.

\section{Discussions}
\subsection{Effect of shear and epicyclic motion}



\begin{figure*}
  \centering

  \includegraphics[width=\textwidth]{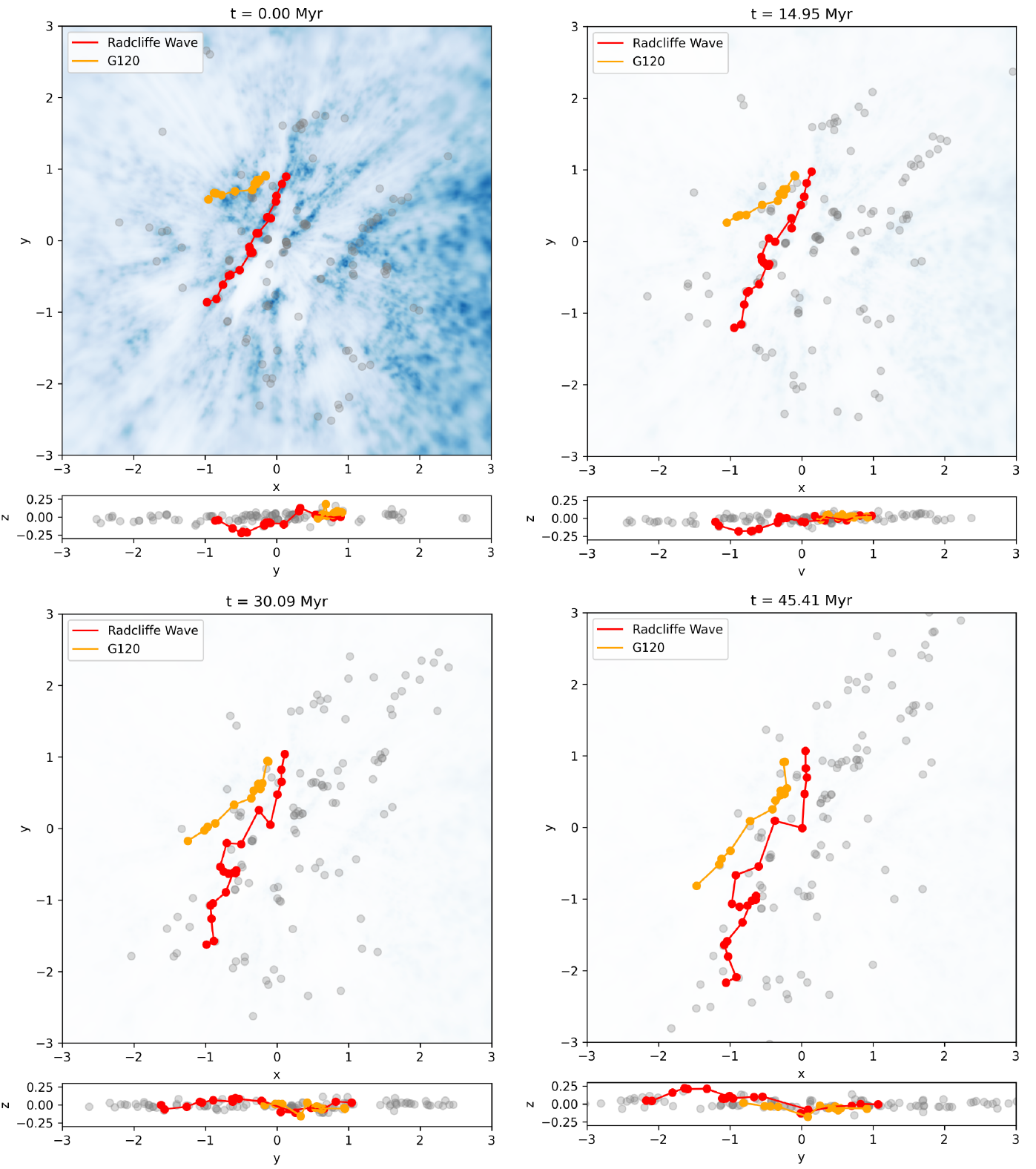}
  \caption{{\bf 3D evolution of the local interstellar gas}. We simulate the evolution  from $t=0$ (now) to $t=45 \;\rm Myr$, where we use the clouds (GMC-YSO complexes  \cite{2022MNRAS.513..638Z,2024arXiv240202393Z}) as anchoring points. The background image is the distribution of dust in the Solar Neighborhood \cite{2022A&A...661A.147L}.
  The clouds are treated as free-moving particles in the Galactic potential.
 The Sun is at $(0,0)$, and the Galactic center is on the right-hand side, outside of the boxes. The plot is presented in the \emph{Local Co-rotating Frame}  \cite{2022MNRAS.516L..35L}, which is located around the Sun and rotates with an angular velocity $v_{\rm circ}/r_{\rm gal}$ where $v_{\rm circ}$ is the circular velocity at the Solar Neighborhood, and $r_{\rm gal}$ is the distance to the Sun.  
  We have labeled different structures using different symbols. These structures include the G120 complex \cite{2022MNRAS.516L..35L}, the Radcliffe wave  \cite{2020Natur.578..237A}. The G120 complex, a present-day gas complex, will evolve into a Radcliffe-wave-like filament, and the Radcliffe wave will be stretched by shear. We also observe encounters between different structures. The full animation is available at  \url{see-attachment-link-to-be-inserted-later}. \label{fig:1}}
  \label{fig:my_figure}
\end{figure*}


From Fig. \ref{fig:1}, one can observe the effect of shear and epicyclic motion on the gas.
 Already at $t= 45 \;\rm Myr$, where the Radcliffe wave has completed half of a
 vertical osculation, it is
 stretched to almost twice its current length by shear, and is out of its current sinusoidal shape. 
 The G120 complex is also significantly
 stretched. We also observe encounters between these Galactic-scale
 structures.

 Shear has a strong effect on evolution for two reasons: First, shear is effective towards large structures such as the Radcliffe where the gravitational acceleration at different parts of its body differs. Second, vertical oscillation is a self-repeating motion, and the effect of shear is \emph{cumulative} such that it can lead to a continuous stretch of the structure. As a result, including shear leads to a significantly different
 picture, where a fast, shear-induced evolution in the disk plane makes the picture of a Wave that oscillates only vertically unfavorable.

  
  \subsection{Implications for mass density estimates}
  The Nature paper further concluded that they 
  
  \begin{quote}
    ``derive its (the Radcliffe wave) motion independently of the local Galactic mass distribution, and
  directly measure local properties of the Galactic potential as well as the Sun's vertical oscillation period.''
  \end{quote}
  However, to achieve this, they needed to describe the evolution of the Wave both in the Galactic disk and in the vertical direction. To achieve the former, they made this assumption that they
  \begin{quote}
  ``take into account Galactic and differential rotation and allow the Wave to
    have an additional overall 2D velocity component, meaning that we permit the structure to move as \emph{a solid
    body} through the Galactic plane with a fixed x- and y-velocity''
  \end{quote}
  Since the Wave is getting stretched inside the disk plane at a time of a few Myr -- a timescale that is critically short compared to that of their proposed oscillation, the assumption that the Wave ``moves as a solid body through the Galactic plane'' is incorrect, and their constraints on the mass density distribution need to be revised.

  \section{Outlook}
  The Nature \cite{2024arXiv240212596K} paper treated the Radcliffe wave as a solid body in the Galactic plane, modeled its evolution, and proposed a picture of the  Radcliffe wave dominated by vertical oscillation.

  We incorporate this vertical oscillation into our 2D kinematic simulation and perform the first 3D forecast simulation of the local interstellar gas. We demonstrate that the picture of a vertically oscillating Radcliffe wave is oversimplified, 
  since shear can change its structure inside the disk plane significantly. 
 
   Our data-driven approach \cite{2022MNRAS.516L..35L} provides a vivid and exciting picture  where structures like the
  Radcliffe waves get created, stretched, merged, and eventually dispersed by the
  Galactic motion. The fact that we can perform such evolution forecasts makes the  Solar Neighborhood an ideal test ground for understanding the physical processes controlling the evolution of interstellar gas in the universe.

\section*{Acknowledgements}



We thank numerous colleagues for their encouragements to write this letter.
GXL acknowledges support from NSFC grant No. 12273032 and
12033005. JXZ is partially supported by the
Post-graduate Research and Innovation Project of Yunnan University
(No. 2019236) and the China Scholarship Council (CSC). BQC is supported by the National Key R\&D Program of China No.
2019YFA0405500, National Natural Science Foundation of China
12173034, 11803029 and 11833006, and the science research grants
from the China Manned Space Project with NO. CMS-CSST-2021-
A09, CMS-CSST-2021-A08 and CMS-CSST-2021-B03.


\bibliography{paper}{}
\bibliographystyle{aasjournal}



\end{document}